\documentclass[useAMS,usenatbib]{mn2e}
\usepackage{times,psfig}
\usepackage{graphics}

\newcommand{\msol}{\rmn{M_{\sun}}}

\title[Mass transfer in tidally unstable binaries]{Mass transfer in tidally unstable compact binaries}
\author[M. R. Truss]{M. R. Truss\\
School of Physics \& Astronomy, University of St Andrews, North Haugh, St Andrews, Fife, KY16 9SS, UK}

\begin{document}

\date{Accepted ~~~ Received ~~~}

\pagerange{\pageref{firstpage}--\pageref{lastpage}} \pubyear{2003}

\maketitle

\label{firstpage}

\begin{abstract}
The 2001 outburst of WZ Sagittae has shown the most compelling evidence yet for an enhancement of the mass 
transfer rate from the donor star during a dwarf nova outburst in the form of hot-spot brightening. I show that even in 
this extreme case, the brightening can be attributed to tidal heating near the interaction point of an accretion stream 
with the expanding edge of an eccentric accretion disc, with no need at all for an increase in the mass transfer rate.
Furthermore, I confirm previous suggestions that an increase in mass transfer rate through the stream damps any
eccentricity in an accretion disc and suppresses the appearance of superhumps, in contradiction to observations.
Tidal heating is expected to be most significant in systems with small mass ratios. It follows that systems like WZ
Sagittae - which has a tiny mass ratio - are those most likely to show a brightening in the hot-spot region.
\end{abstract}
\begin{keywords}
accretion, accretion discs - binaries:close - novae, cataclysmic variables - stars, individual: WZ Sagittae.
\end{keywords}
\section{Introduction}
Dwarf novae are cataclysmic variable stars that are observed to undergo bright outbursts lasting a few days. 
These outbursts are separated by weeks to months of dim quiescence. At the short orbital period end of the dwarf 
nova distribution ($P_{\rm orb} < 2.2$ h), we find the SU UMa stars, which also show superoutbursts. These are 
longer outbursts lasting a couple of weeks or more, with the additional presence of superhumps, a periodic modulation in 
the $V$ band light curve that repeats on a time-scale very close to the orbital period. \citet{war} gives a comprehensive 
review of dwarf novae.

The outbursts of dwarf novae have been explained in terms of an instability associated with the gas circulating in 
the accretion disc. \citet{hos} realised that there is a rapid change in opacity when hydrogen becomes ionised near 
$6\,500\,\rmn{K}$. If the temperature is not higher than this everywhere in the disc, a cyclic behaviour can be 
established between two thermally stable states: a hot, ionised outburst state and a cool, neutral quiescent state. 
It was realised almost immediately that such a thermal change needs to be accompanied by a change in the gas 
viscosity to give the change in angular momentum transport that is required to power the observed accretion rates 
in dwarf novae. The best candidate for this viscosity remains magnetohydrodynamic turbulence, as suggested by 
\citet{bal}. Historically, an alternative outburst theory was proposed involving an instability in the 
mass-transfer rate from a convectively unstable envelope of a heated donor star \citep{pac}. It was the disc 
instability theory that prevailed due to better compatibility with observations, which in nearly all cases have 
showed no evidence for an enhanced mass transfer rate from the companion star. However, a re-evaluation of 
some of the observations \citep{sma95} and evidence from eclipses during the latest outburst of WZ 
Sagittae \citep{pat02} has led to renewed interest in the possibility of enhanced mass transfer. 

In a recent paper, \citet{osa03} reviewed the arguments for and against enhanced mass transfer. They suggested that 
the eclipses observed for WZ Sagittae could be eclipses of the superhump light source and not the hot spot. 
The case of WZ Sagittae is discussed in more detail in Section 2 below, as well as the evidence from existing numerical 
work not discussed by Osaki \& Meyer.

\citet{osa85} originally proposed a model in which superoutbursts were caused by an enhanced mass-transfer 
rate from an irradiated secondary star.  However, numerical experiments by \citet{whi} showed that discs in a 
Roche potential with $q < 0.3$ could spread to a radius at which the material at the edge of the disc becomes 3:1 
resonant with the tidal field of the secondary star. At this point the disc becomes unstable and the gaseous orbits 
become eccentric. Since superoutbursts are only observed in systems with low mass ratios, it seems likely that 
such a tidal instability plays a role in their development. Soon after Whitehurst's experiments, Osaki formulated the 
thermal-tidal instability (TTI) model \citep{osa89}.  The mass transfer rate from the donor remains constant in the 
TTI model, but with every passing normal outburst the accretion disc expands until it reaches the 3:1 resonant 
radius. At this point, Osaki suggested that the rate of angular momentum loss becomes more rapid, leading to the 
accretion of a large amount of mass, and a superoutburst. The TTI picture was broadly supported by the first 
two-dimensional (2D) hydrodynamic models of superoutbursts \citep{tru01}, which confirmed that while the 
outburst is initiated by disc instability in the same way as a normal outburst, the tidally-enhanced energy 
dissipation prolongs the outburst, explaining the difference in duration between normal and superoutbursts. A 
significant eccentricity was generated and superhumps did indeed develop on contact with the 3:1 radius.

In the next section I introduce the evidence for enhanced mass transfer in SU UMa stars from observations of hot 
spot brightness and balance this with a discussion of existing theoretical studies of mass transfer in these systems. In 
Section 3 I present the results of new, three-dimensional (3D) simulations of tidally unstable 
accretion discs. These show that tidal heating produces a significant brightening near the stream-impact region, even 
with no increase in the mass transfer rate through the stream. This means that a simple energetic model for hot-spot 
luminosity is inadequate for a low-$q$ system in outburst, as the luminosity in this region is dominated by tidal effects.
The results therefore lend weight to the arguments of \citet{osa03} that the brightening observed near the 
stream-impact region in some SU UMa stars is not due to the hot-spot, but is due to the tidal heating of the outer parts 
of the disc.

\section{Enhanced mass transfer: observation versus simulation}

Until the 2001 superoutburst of WZ Sagittae, little evidence for enhanced mass transfer during dwarf nova outbursts
had been found. A notable exception is that of VW Hydri; \citet{vog} found that the amplitude of a hump in the light 
curve increased during normal outbursts occurring within 40 days of a subsequent superoutburst. This was interpreted
as brightening of the hot spot in response to an increase in mass transfer rate from the donor star. \citet{sma95} 
used similar arguments for an increase in mass transfer rate by a factor of two in outbursts of U Geminorum and 
Z Chamaeleontis, employing the relationship
\begin{equation}
   L_{\rmn{spot}} = \frac{1}{2} \dot M_2 \left( {\bf v}_{\rmn{K}} - {\bf v}_{\rmn{b}}\right)^2
\label{lspot}
\end{equation}
where $\dot M_2$ is the mass transfer rate from the donor star, ${\bf v}_{\rmn K}$ is the Keplerian velocity of the
gas circulating in the accretion disc and ${\bf v}_{\rmn b}$ is the ballistic velocity of the in-falling gas stream. 

WZ Sagittae is the brightest known dwarf nova in the $V$-band, with $V \sim 8$ at supermaximum. It is unusual in 
only showing superoutbursts, and in its extremely long recurrence time: the 2001 outburst occurred a mere 23 
years after a 33 year cycle of previous events in 1913, 1946 and 1978. It has a high-inclination ($i\sim75^{\circ}$) 
and an orbital period of 81.6 minutes \citep{krz}. The secondary star is of unusually low mass, with $M_{\rmn{2}} =
0.045 \pm 0.003\,\msol$ and $q \sim 0.06$ \citep{ski}. The 2001 outburst was observed in great detail, revealing 
many layers of variability. Superhumps appeared in the light curve 13 days after the onset of the outburst, 
reaching maximum amplitude 24 hours later. They persisted for 90 days, well into the decline to quiescence, but 
were preceded by a different modulation on the orbital period for the first 12 days. \citet{pat02} found that the 
depth of eclipses in the light curve increased rapidly 6 days after the appearance of superhumps, and then
gradually decreased. This change was mirrored by the height of the orbital hump in the light curve. The inclination of
WZ Sge is not high enough to make eclipses of the white dwarf itself observable, so the eclipses in the light curve are
assumed to be associated with the hot-spot. Patterson et al. concluded that the hot-spot brightened by a factor 60, 
which is suggestive of a commensurate increase in the mass-transfer rate from the secondary star.

I will not discuss the theoretical aspects relating to the response of an irradiated star in terms of mass transfer 
rate here; most recently these arguments have been made by \citet{osa03}. Instead, I will discuss briefly the 
application of analytic methods and hydrodynamic simulations to the problem of stream-disc interactions. \citet{lub} 
investigated the response of an eccentric accretion disc to an increase in the mass transfer rate through a stream.
He found that rather than pumping the eccentricity up further, the addition of low specific angular momentum
material at a faster rate tended to circularise the gas orbits. Later smoothed particle hydrodynamics simulations of
dwarf novae with mass ratios near the border of tidal stability confirmed that superhumps appeared when the 
mass transfer rate was reduced from a rather high value, but disappeared when the rate was restored 
\citep{mur00}. These theoretical results cannot explain the persistence of the superhumps in WZ Sge with an 
enhanced mass transfer rate. One could circumvent this problem if the superoutburst itself was initiated by a 
burst of mass transfer and the disc radius expanded from a value well inside the 3:1 radius to a radius beyond it. 
Unfortunately, this argument fails for WZ Sge. The observed increase in the eclipse depth occurs six days after 
the appearance of superhumps, so the disc radius has already expanded far beyond the 3:1 radius by the time the 
hot-spot region is observed to brighten.

Recently, \citet{fou} have used a 2D hydrodynamic scheme to model the evolution of spectral line profiles 
generated by an eccentric accretion disc in a binary with $q$ = 0.1. They found significant changes in the profile 
over the precession cycle of the disc and were able to distinguish between the individual components from the 
stream-impact region and the spiral arms. Both components contribute to the superhumps seen in the light curve.

\section{Results}

In this section I present the results from new, three-dimensional smoothed particle hydrodynamics (SPH)
simulations of the accretion disc and stream in WZ Sge. The simulations were performed on the UK Astrophysical 
Fluids Facility (UKAFF) supercomputer during time awarded in Spring/Summer 2004.
\subsection{Numerical Method}
I use a parallelised version of a 3D SPH code optimised for the simulation of accretion discs in compact binary stars.
It is an evolution of the original code developed by \citet{mur96}. In the SPH method, the fluid equations are solved for 
each of a set of moving points which we call particles. Local physical quantities are calculated by means of a weighted
sum over all the nearby particles within a set distance. This distance, called the smoothing length, $h$, is tuned to
the local density, making the method extremely adaptable in resolving large density contrasts. A review of the SPH 
method can be found in \citet{mon}. In this work, $h$ is calculated for each particle such that a minimum of 50 neighbour 
particles is maintained regardless of the local density. Here, a total particle number of order 400,000 allows a resolved 
lengthscale $h < 0.01R_{\rmn{disc}}$ over most of the accretion disc.

The simulation proceeds by the continuous injection of particles from the inner Lagrangian ($L_1$) point. The gas 
dynamics of the particles is calculated in the full 3D Roche potential of a binary system, with particles being rejected 
from the simulation if they pass through an inner boundary of the disc at a radius $R < 0.05a$ where $a$ is
the separation of the two stars, if they intersect the Roche lobe of the secondary star or if they achieve the escape
velocity at a radius larger than a. The viscosity used in the SPH calculation is a refinement of that described by \citet{mur96} 
to mimic an $\alpha$-viscosity:
\begin{equation}
   \nu = \alpha c_{\rmn{s}} H = \alpha c_{\rmn{s}}^2/\Omega.
\end{equation}
The lengthscale is chosen to be the scale height $H$ rather than the smoothing length, giving closer compatibility 
with the \citet{sha} prescription.

No viscous calculation is performed for the particles in the stream, which move on a purely ballistic orbit. The equation 
of state of all the particles is isothermal, so the rate of energy generation in the disc is purely from the viscous dissipation 
induced by the interaction of the particle orbits. Therefore, the effects of radiative processes are neglected, making this a 
simple calculation of the total bolometric luminosity due to viscous dissipation at each point in the disc.   

\subsection{Simulations}

I use parameters representative of WZ Sge: $q = 0.07$, $P_{\rmn{orb}} = 0.056 \ \rmn{days}$, $-\dot M_2 = 3 \times 
10^{-11} \ \rmn{M_\odot \ yr^{-1}}$ \citep[from a quiescent hot-spot brightness]{sma93}. The thermal parameters of
the particles in the disc are chosen as a conservative estimate for a CV disc in outburst, that is at the lower end of their
probable range: $\alpha = 0.1$, $c_{\rmn{s}} = 
0.04 \ a\Omega_{\rmn{orb}}$, which assuming a mean molecular mass $\mu =0.6$ corresponds to a mid-plane 
temperature $T_{\rmn{c}} = 42,500 \ \rmn{K}$. The initial conditions for the 3D simulation are generated in the same 
way as in \citet{mur02}, where a 3D disc is replicated from a particle distribution derived from an initial 2D simulation. 
The conditions are then relaxed to a steady-state before the 3D simulation proper. The disc is built up using a constant
mass transfer rate of $3 \times 10^{-11} \rmn{M_{\odot} \, yr^{-1}}$. The 3D runs start at a point after the outer edge 
of the disc has crossed the 3:1 radius and superhumps have developed. Three runs are then performed with initial
conditions that are identical except for the rate of mass injection. Run A continues with the same mass transfer rate.
Run B proceeds with a rate five times lower, run C with a rate five times higher. The number of particles at the start
of each run was $N = 428685$.
\begin{figure}
  \psfig{file=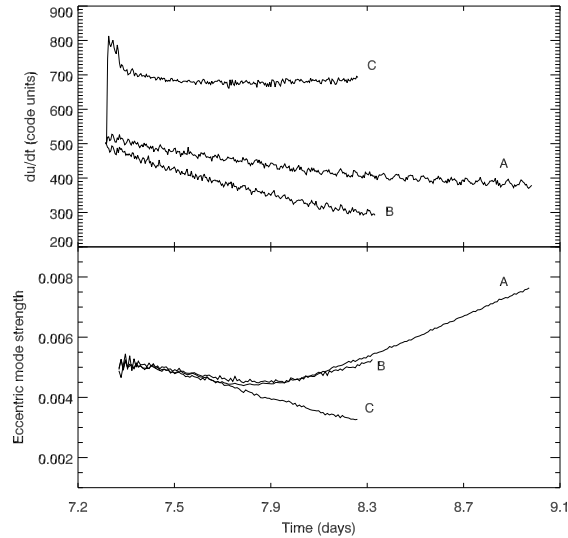,width=8cm}
  \caption{Viscous energy dissipation rate per unit mass and eccentric mode strength during the three simulations. A: 
$-\dot M_2 = 3 \times 10^{-11} \rmn{M_{\odot} \ yr^{-1}}$, B: $-\dot M_2 = 6 \times 10^{-12} \rmn{M_{\odot} \ yr^{-1}}$,
C: $-\dot M_2 = 1.5 \times 10^{-10} \rmn{M_{\odot} \ yr^{-1}}$.}
  \label{fig1}
\end{figure}

Fig.\ref{fig1} shows the evolution of the viscous energy dissipation rate and the $(k,l)=(1,0)$ eccentric mode,
calculated using the method described in Section 3.3 of \citet[hereafter TMW]{tru01}. The superhumps are clearly visible in the run with
the unchanged mass transfer rate (A) and in the run with the reduced rate (B). However, they disappear when the 
rate is increased in run C, with the eccentricity falling throughout the simulation as the disc edge is driven away from 
the 3:1 radius. The difference in length of the simulations is due to the finite supercomputer time.
\begin{figure*}
  \psfig{file=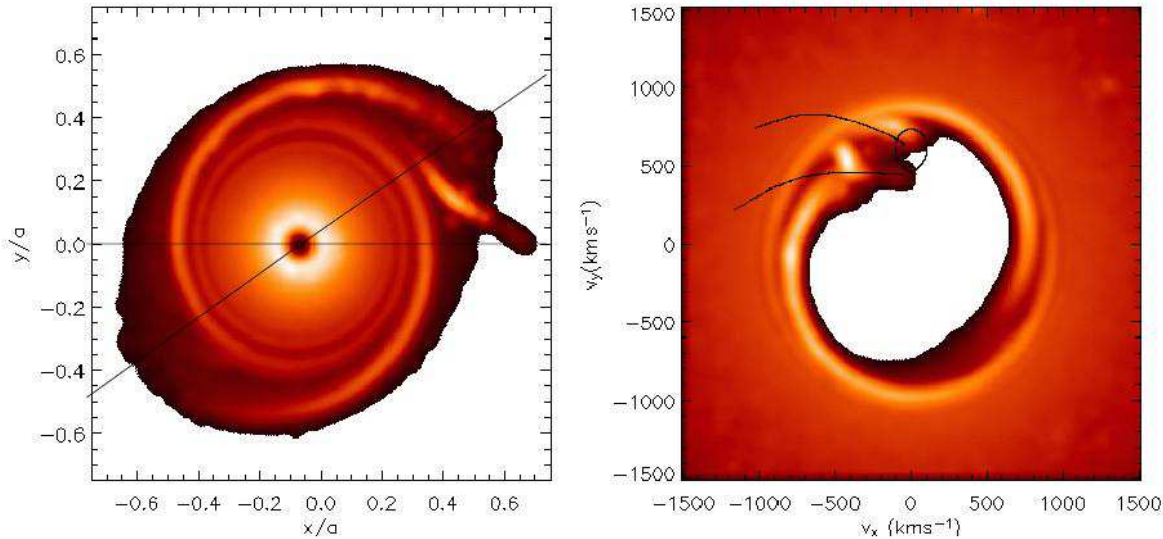,width=16cm}
  \caption{Map of energy dissipation rate in the simulation with $-\dot M_2 = 3 \times 10^{-11} \rmn{M_{\odot}\,yr^{-1}}$
and corresponding Doppler velocity map. The straight lines on the left define the sectors that are considered in 
generating the radial profiles in Fig. \ref{fig3}. The Roche lobe of the secondary star and velocities of the ballistic 
stream and its Keplerian equivalent are plotted on the right for $q = 0.07$.} 
  \label{fig2}
\end{figure*}

\begin{figure}
  \psfig{file=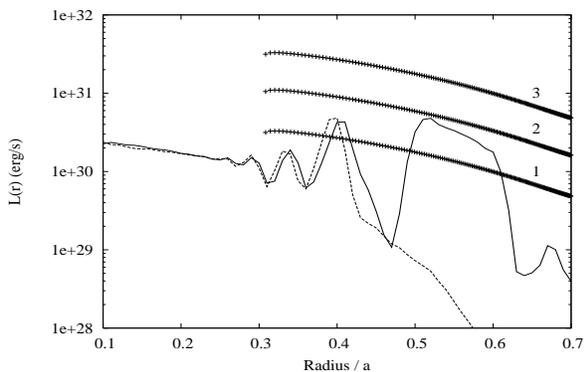,width=8cm,height=50mm,angle=-90}
  \caption{Radial profile of luminosity per radial bin of width 0.01a for the two sectors in Fig. \ref{fig2}.
The solid line has been smoothed over the sector that includes the hot-spot, the dashed line is the equivalent for the
diametrically opposite sector. The numbered curves are the expected total bolometric hot-spot luminosity
calculated from equation \ref{lspot} for $-\dot M_2 = 3 \times 10^{-11}$ (1), $10^{-10}$ (2) and  $3 \times 10^{-10} 
\rmn{M_{\odot}\,yr^{-1}}$ (3).}
  \label{fig3}
\end{figure}

Fig.\ref{fig2} shows a map of the energy dissipation rate in the disc and a corresponding Doppler velocity map at 
time $t = 8$ days for run A. The hot-spot region and tidally-induced spiral structure are clearly evident, as are velocity
components consistent with gas on ballistic and Keplerian trajectories. There is also a non-Keplerian component in the 
velocity map, similar to that noticed by \citet{fou}. The radial luminosity profile of the disc is presented in Fig.\ref{fig3} 
for the two sectors shown in the dissipation map. This has been calculated by dividing the disc up into
annular bins of width $0.01a$ and averaging the luminosity (dissipation rate per unit area multiplied by the area of each 
bin) in the azimuthal direction. Also plotted is the expected total bolometric hot-spot luminosity calculated from 
equation \ref{lspot} for three different mass transfer rates. Even with this small emitting area per radial bin, the 
luminosity in the hot spot region is well above that expected for $-\dot M_2 = 3 \times 10^{-11} \rmn{M_{\odot}\,yr^{-1}}$.
If one assumes that the radial extent of the hot-spot covers the region of tidally-enhanced dissipation between $R=0.5a$
and $R=0.6a$, the total 'hot-spot' luminosity is nearly an order of magnitude higher still.

A feature that becomes apparent from the 3D simulations is the magnitude of the eccentricity, which is far
smaller than has been found in previous 2D work. The mode strength in Fig.\ref{fig1} is of the order a few times 
$10^{-3}$, in comparison with $\sim 0.5$ for 2D superoutburst simulations of Z Cha (TMW). As a result, the pattern of
dissipation shown in Fig. \ref{fig2} does not change appreciably over the course of the simulation, and the dissipation remains
enhanced in the region of the hotspot throughout. However, the Fig.\ref{fig1} shows that the eccentricity of the 3D discs in
runs A and B are increasing at the end of the simulations. It remains unclear whether this trend would continue if the 
simulations were allowed to run indefinitely, although the growth rate of the eccentric mode in 3D does not suggest that the
eccentricity will reach the levels achieved in the 2D case over a sensible time-scale for a dwarf nova outburst. 

Despite the low eccentricity, a clear superhump signal is still seen and the tidal interaction still has a significant influence on 
the luminosity generated in the hot-spot region. Of course, if the eccentricity were higher the interaction would be even 
stronger. We would also expect to see more modulation in the dissipation map over a precession cycle, as discussed by 
\citet{fou}. 

\begin{figure*}
  \psfig{file=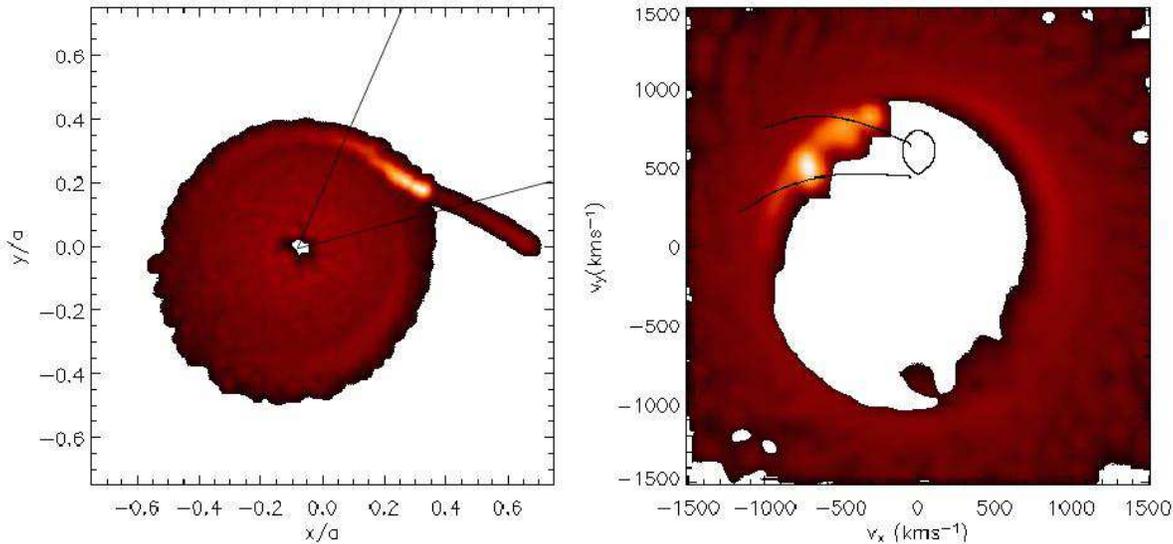,width=16cm}
  \caption{As Fig.\ref{fig2}, but for a smaller disc with $R_{\rmn{disc}} <  R_{\rmn{3:1}}$. The ballistic and Keplerian
components are much more clearly defined in the Doppler map and there are only very weak spiral features. The straight
lines on the left hand side define the sector of the disc used to calculate the luminosity profile in Fig.\ref{fig5}.}
  \label{fig4}
\end{figure*}

\begin{figure}
  \psfig{file=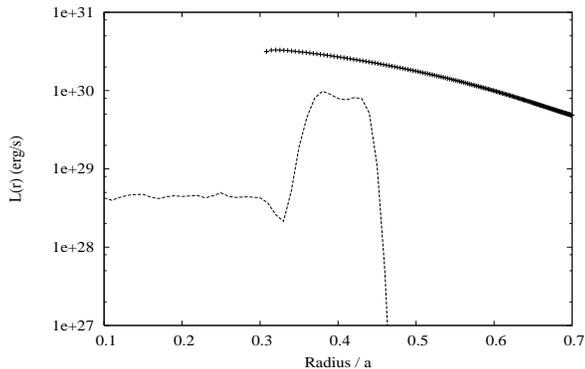,width=8cm,height=50mm,angle=-90}
  \caption{As Fig.\ref{fig3}, but for a smaller disc.The integrated hot spot luminosity is consistent with the
curve calculated from equation \ref{lspot} with $-\dot M_2 = 3 \times 10^{-11} \rmn{M_{\odot}\,yr^{-1}}$.}
  \label{fig5}
\end{figure}

As a final check, a short simulation was performed for a disc not subject to the tidal instability, with an outer radius
well inside the 3:1 resonance. The results are shown in Figs. \ref{fig4} and \ref{fig5}. In this case, the structure of the 
hot spot is exactly as expected. There are clearly defined components of emission at the ballistic and Keplerian velocities
and very little else, only a very weak spiral structure. The integrated luminosity of the hot-spot is now consistent with
equation \ref{lspot} for the mass transfer rate that has been measured in quiescence, and that is used in the simulation:
$-\dot M_2 = 3 \times 10^{-11} \rmn{M_{\odot}\,yr^{-1}}$.

\section{Discussion}
I have shown that tidal heating accounts for a significant fraction of the energy dissipated near the stream-impact 
region in an eccentric accretion disc. The tidal heating component dominates that due to the impact of the
ballistic stream, producing an apparent brightening near the hot-spot. No increase in the mass transfer rate from 
the secondary star is required to produce this brightening. Therefore, even for a disc with low eccentricity the simple 
model for the luminosity of the hot-spot given in equation \ref{lspot} is inadequate to estimate the mass transfer
rate from the donor. Of course, this equation is not incorrect: but for a small correction due to the slightly non-
Keplerian velocity of the orbiting gas in the disc, the luminosity of the impact alone will be given correctly. It is
just that this luminosity is swamped by the tidal heating component. \citet{osa03} 
suggested that it is the eclipse of the superhump light source (dominated by the tides) and not the hot-spot that 
has been observed in WZ Sge. This work supports this view, with the caveat that it could well be that the eclipse
is that of the hot-spot, the luminosity in this region is simply being dominated by the tides. For a disc with zero 
eccentricity, such as a quiescent disc with an outer radius much smaller than the 3:1 radius, the expression in 
equation \ref{lspot} remains a good diagnostic of the mass transfer rate.

This brightening of the hot spot region will be observed in systems with mass ratio $q = M_2/M_1 < 0.3$, where tidal 
instability becomes important. It may seem intuitive to take this assertion a step further; that the lower the mass ratio, 
the stronger the tidal heating and the brighter the hot-spot. However, it should be stressed that to date no systematic 
studies of the variation in strength of the tidal heating component with mass ratio have been carried out. 

In terms of the sequence of events during an outburst of WZ Sge, the picture that has emerged from this work 
would be as follows. I assume that the outburst is initiated in the normal way by the disc instability, with the disc
spreading past the 3:1 radius some time later. The growth rate of the eccentricity must be very fast as the 
superhumps were observed to reach maximum amplitude within 24 hours of appearing in the light curve 
\citep{pat02}. The brightening effect due to the tides will be first observed up to one precession cycle later when 
the apastron of the disc passes $L_1$. This is consistent with the delay of a few days that was observed between
the appearance of the superhumps and the brightening event. As the disc empties and the outer radius finally shrinks 
inside the 3:1 radius, the effect will diminish over the course of the outburst until the brightness of the hot-spot 
becomes consistent with the value given by equation \ref{lspot} for the mass transfer rate measured in quiescence. 

I have confirmed with a 3D model the analytic prediction of \citet{lub} and the 2D prediction of 
\citet{mur00} that an increase in mass transfer rate from the donor drives the disc away from the 3:1 resonance
and suppresses the appearance of superhumps. This is in contradiction to what has been observed in WZ Sge.
There are interesting differences between the 3D results presented here and previous 2D work. In particular, while
eccentricity and superhumps develop in the same way in 2D and 3D, it is much more difficult to generate high 
eccentricities in 3D for reasonable values of sound speed and viscosity parameter. Indeed, the eccentricity of
$e \geq 0.3$ estimated for WZ Sge by \citet{pat02} is not found in these 3D simulations. However, the tidal heating 
effect is still significant even with a small eccentricity. If future observations can reveal with accuracy the exact 
shape of the disc and $e$ is found to be as high as 0.3, we will need to consider how this can be achieved. A 
possible solution may lie in a magnetic field anchored on the rotating white dwarf, which could have a 
propelloring effect on the gas \citep{liv,las} that keeps the outer edge of the disc past the 3:1 radius and allows
a large eccentricity to develop. I will investigate this numerically in a forthcoming paper.

As an aside to these issues, it is interesting to note that an eccentric disc has no well-defined single outer radius. 
Indeed, if we assume the disc is an ellipse with an eccentricity $e$, the ratio of the semi-major to semi-minor axes is
\begin{equation}
   \frac{r_{\rmn{+}}}{r_{\rmn{-}}} = \frac{1+e}{1-e}.
\end{equation}
So, for $e \sim 0.3$, there is nearly a factor of 2 variation in apparent radius over a single
precession cycle. This should always be borne in mind when inferring accretion disc radii from observations in
tidally unstable systems, especially given that the precession period can be several days in comparison to the
orbital or superhump periods, which are usually shorter than a couple of hours.

In conclusion, it is apparent from this work that the observations of hot spot brightnesses during dwarf nova
outbursts can be attributed to the interaction of a constant-$\dot M$ accretion stream with the edge of an
eccentric accretion disc. Future work on the theoretical side should take into account the effects of radiative transfer in 
the spot region.
This will be a challenge, and has not been considered in this work, however it is expected to have an impact on the
predicted brightnesses. In particular, the amount of stream overflow above and below the rim of the disc is sensitive to 
the radiative cooling rate in the hot-spot region \citep{arm,kun}. This will have observable consequences on the shape of the measured eclipses and the appearance of the orbital 
hump in the light curve. From an observational point of view, it is desirable to discover the exact dimensions
and geometry of a precessing disc during outburst. This will be a robust test of the TTI model for superoutbursts, 
which relies on the disc edge encountering the 3:1 resonance and the mass transfer rate remaining fixed.

\section*{Acknowledgments}
I acknowledge a PPARC Postdoctoral Research Fellowship at the University of St Andrews. 
The simulations were performed on the UK Astrophysical Fluids Facility (UKAFF). I thank Keith Horne, Mario Livio and 
Danny Steeghs for helpful discussions, and also Theo Truss, who kindly waited until the day after this paper was completed 
before he was born.

\label{lastpage}


\begin{thebibliography}{99}
\bibitem[\protect\citeauthoryear{Armitage \& Livio}{1998}]{arm} Armitage P., Livio M., 1998, ApJ, 493, 898
\bibitem[\protect\citeauthoryear{Balbus \& Hawley}{1991}]{bal} Balbus S.A., Hawley J.F., 1991, ApJ, 376, 214  
\bibitem[\protect\citeauthoryear{Foulkes et al.}{2004}]{fou} Foulkes S.B., Haswell C.A., Murray J.R., Rolfe D.J., 2004, MNRAS, 349, 1179
\bibitem[\protect\citeauthoryear{Hoshi}{1979}]{hos} Hoshi R., 1979, Prog. Theor. Phys., 61, 1307
\bibitem[\protect\citeauthoryear{Krzeminski}{1962}]{krz} Krzeminski W., 1962, PASP, 74, 66
\bibitem[\protect\citeauthoryear{Kunze, Speith \& Hessman}{2000}]{kun} Kunze S., Speith R., Hessman F.V., 2000, MNRAS, 322, 499
\bibitem[\protect\citeauthoryear{Lasota, Kuulkers \& Charles}{1999}]{las} Lasota J.-P., Kuulkers E., Charles P.A., 1999, MNRAS, 305, 473
\bibitem[\protect\citeauthoryear{Livio \& Pringle}{1994}]{liv} Livio M., Pringle J.E., 1994, MNRAS, 259, 23 
\bibitem[\protect\citeauthoryear{Lubow}{1994}]{lub} Lubow S.H., 1994, ApJ, 432, 224
\bibitem[\protect\citeauthoryear{Monaghan}{1992}]{mon} Monaghan J.J., 1992, ARA\&A, 30, 543
\bibitem[\protect\citeauthoryear{Murray}{1996}]{mur96} Murray J.R., 1996, MNRAS, 279, 402
\bibitem[\protect\citeauthoryear{Murray, Warner \& Wickramasinghe}{2000}]{mur00} Murray J.R., Warner B., Wickramasinghe D.T., 2000, MNRAS, 315, 707
\bibitem[\protect\citeauthoryear{Murray et al.}{2002}]{mur02} Murray J.R., Chakrabarty D., Wynn G.A., Kramer L., 2002, MNRAS, 335, 247
\bibitem[\protect\citeauthoryear{Osaki}{1985}]{osa85} Osaki Y., 1985, A\&A, 144, 369
\bibitem[\protect\citeauthoryear{Osaki}{1989}]{osa89} Osaki Y., 1989, PASP, 41, 1005
\bibitem[\protect\citeauthoryear{Osaki \& Meyer}{2003}]{osa03} Osaki Y., Meyer F., 2003, A\&A, 401, 325
\bibitem[\protect\citeauthoryear{Paczy\'nski, Ziolkowski \& Zytkow}{1969}]{pac} Paczy\'nski B., Ziolkowski J., Zytkow A., 1969, in Mass Loss from Stars, ed. Hack M., Reidel, Dordrecht
\bibitem[\protect\citeauthoryear{Patterson et al.}{2002}]{pat02} Patterson J. et al., 2002, PASP, 114, 721
\bibitem[\protect\citeauthoryear{Shakura \& Sunyaev}{1973}]{sha} Shakura N.I., Sunyaev R.A., 1973, A\&A, 24, 337
\bibitem[\protect\citeauthoryear{Skidmore et al.}{2002}]{ski} Skidmore W., Wynn G.A., Leach R., Jameson R.F., 2002, MNRAS, 336, 1223
\bibitem[\protect\citeauthoryear{Smak}{1993}]{sma93} Smak J., 1993, Acta Astr., 43, 101
\bibitem[\protect\citeauthoryear{Smak}{1995}]{sma95} Smak J., 1995, Acta.Astr., 45, 355
\bibitem[\protect\citeauthoryear{Truss, Murray \& Wynn}{2001}]{tru01} Truss M.R., Murray J.R., Wynn G.A., 2001, MNRAS, 324, L1
\bibitem[\protect\citeauthoryear{Vogt}{1983}]{vog} Vogt N., 1983, A\&A, 118, 95
\bibitem[\protect\citeauthoryear{Warner}{1995}]{war} Warner B., 1995, Cataclysmic Variable Stars, Cambridge University Press
\bibitem[\protect\citeauthoryear{Whitehurst}{1988}]{whi} Whitehurst R., 1988, MNRAS, 232, 35
\end{thebibliography}
\end{document}